\documentclass[preprint]{sigplanconf}

\usepackage{amsmath}
\usepackage{graphicx}
\usepackage{float}
\usepackage{textcomp}
\usepackage[pdfborder={0 0 0}]{hyperref}

\begin{document}

\conferenceinfo{Memory Systems Performance and Correctness 2011}{June 5th 2011, San Jose, U.S.A.} 
\copyrightyear{2010-2011} 
\copyrightdata{Niall Douglas} 

\titlebanner{Draft 1}        
\preprintfooter{Draft 1}   

\title{User Mode Memory Page Management}
\subtitle{An old idea applied anew to the memory wall problem}

\authorinfo{Mr. Niall Douglas BSc MA MBS MCollT}
           {ned Productions IT Consulting}
           {\url{http://www.nedproductions.biz/}}
\hypersetup{pdftitle={User Mode Memory Page Management: An old idea applied anew to the memory wall problem}, pdfauthor={Niall Douglas}, pdfsubject={Memory Systems Performance and Correctness 2011}, pdfkeywords={Performance, System Balance, MMU, mmap, sbrk, malloc, realloc, free, O(1), virtualization, N1527, C1X, kernel, memory wall}}

\maketitle

\begin{abstract}
It is often said that one of the biggest limitations on computer performance is memory bandwidth (i.e.``the memory wall problem"). In this position paper, I argue that if historical trends in computing evolution (where growth in available capacity is exponential and reduction in its access latencies is linear) continue as they have, then this view is wrong -- in fact we ought to be concentrating on \emph{reducing whole system memory access latencies} wherever possible, and by ``whole system'' I mean that we ought to look at how software can be unnecessarily wasteful with memory bandwidth due to legacy design decisions.

To this end I conduct a feasibility study to determine whether we ought to virtualise the MMU for each application process such that it has direct access to its own MMU page tables and the memory allocated to a process is managed exclusively by the process and not the kernel. I find under typical conditions that nearly scale invariant performance to memory allocation size is possible such that hundreds of megabytes of memory can be allocated, relocated, swapped and deallocated in almost the same time as kilobytes (e.g. allocating 8Mb is 10x quicker under this experimental allocator than a conventional allocator, and resizing a 128Kb block to 256Kb block is 4.5x faster). I find that first time page access latencies are improved tenfold; moreover, because the kernel page fault handler is never called, the lack of cache pollution improves whole application memory access latencies increasing performance by up to 2x. Finally, I try binary patching existing applications to use the experimental allocation technique, finding almost universal performance improvements without having to recompile these applications to make better use of the new facilities.

As memory capacities continue to grow exponentially, applications will make ever larger allocations and deallocations which under present page management techniques will incur ever rising bandwidth overheads. The proposed technique removes most of the bandwidth penalties associated with allocating and deallocating large quantities of memory, especially in a multi-core environment. The proposed technique is easy to implement, retrofits gracefully onto existing OS and application designs and I think is worthy of serious consideration by system vendors, especially if combined with a parallelised batch allocator API such as N1527 currently before the ISO C1X programming language committee.
\end{abstract}




\section{Introduction}
There is a lot in the literature about the memory wall problem, and most of it comes from either the hardware perspective \cite{wulf1995hitting, saulsbury1996missing, mckee2004reflections, cristal2005kilo, boncz2008breaking} or from the software and quite tangential perspective of the effects upon performance of memory allocation techniques. As a quick illustration of the software approaches, the Lea allocator, dlmalloc \cite{lea2000memory}, aims for a reusable simplicity of implementation, whereas other allocators have a much more complex implementation which makes use of per-processor heaps, lock-free, cache-line locality and transactional techniques \cite{berger2000hoard, michael2004scalable, hudson2006mcrt}. Many still believe strongly in the use of custom application-specific allocators despite that research has shown many of these implementations to be sub-par \cite{berger2002reconsidering}, whereas others believe that the enhanced type information and metadata available to source compilers allow superior allocators to be implemented at the compilation stage \cite{berger2001composing, udayakumaran2003compiler, boostpool2010}.

Yet there appears to me to be a lack of joined up thinking going on here. The problem is not one of either hardware or software alone, but of \emph{whole application performance} which looks at the \textbf{entire} system including the humans and other systems which use it -- a fundamentally \textbf{non-linear} problem. There is some very recent evidence in the literature that this holistic approach is becoming realised: for example, Hudson (2006) \cite{hudson2006mcrt} and Dice (2010) \cite{dice2010simplifying, Dice2010} looked into how a memory allocator's implementation strategy non-linearly affects whole application performance via cache set associativity effects and cache line localisation effects, with a very wide range of approaches and techniques suggested.

To quote the well known note on hitting the memory wall by Wulf and McKee (1995) \cite{wulf1995hitting}:
\begin{quote}
\emph{``Our prediction of the memory wall is probably wrong too -- but it suggests that we have to start thinking ``out of the box''.
All the techniques that the authors are aware of, including ones we have proposed, provide one-time
boosts to either bandwidth or latency. While these delay the
date of impact, they don't change the fundamentals."} (p. 22).
\end{quote}

Personally I don't think that the fundamentals \emph{are} changeable -- the well established holographic principle\footnote{It even has a Wikipedia page which is not bad at \url{http://en.wikipedia.org/wiki/Holographic_principle}.} clearly shows that maximal entropy in any region scales with its radius \emph{squared} and not cubed as might be expected \cite{bekenstein2003information} i.e. it is the \emph{boundary} to a region which determines its maximum information content, not its volume. Therefore, growth in storage capacity and the speed in accessing that storage will always diverge exponentially for exactly the same reason as why mass -- or organisations, or anything which conserves information entropy -- face decreasing marginal returns for additional investment into extra order.

This is not to suggest that we are inevitably doomed in the long run -- rather that, as Wulf and McKee also suggested, we need to start revisiting previously untouchable assumptions. In particular, we need to identify the ``low hanging fruit'' i.e. those parts of system design which consume or cause the consumption of a lot of memory bandwidth due to legacy design and algorthmic decisions and replace them with functionally equivalent solutions which are more intelligent. This paper suggests one such low hanging fruit -- the use of page faulted virtual memory.

\section{How computing systems currently manage memory}

Virtual memory, as originally proposed by Denning in 1970 \cite{denning1970virtual}, has become so ingrained into how we think of memory that it has become taken for granted. However, back when it was first introduced, virtual memory was \emph{controversial} and for good reason, and I think it worthwhile that we recap exactly why.

Virtual memory has two typical meanings which are often confused: (i) the MMU hardware which maps physical memory pages to appear at arbitrary (virtual) addresses and (ii) the operating system support for using a paging device to store less frequently used memory pages, thus allowing the maximal use of the available physical RAM in storing the most used memory pages. As Denning pointed out in 1970 \cite{denning1970virtual}, again with much more empirical detail in 1980 \cite{denning1980working} and indeed in a reflection upon his career in 1996 \cite{denning1996}, virtual memory had the particular convenience of letting programmers write \emph{as if} the computer had lots of low latency memory. The owner of a particular computer could then trade off between cost of additional physical RAM and execution speed, with the kernel trying its best to configure an optimal \emph{working set} of most frequently used memory pages for a given application.

We take this design for granted nowadays. But think about it: there are several assumptions in this choice of design, and these are:
\begin{enumerate}
\item That lower latency memory capacity (physical RAM) is expensive and therefore scarce compared to other forms of memory.
\item That higher latency memory capacity (magnetic hard drives) is cheaper and therefore more plentiful.
\item That growth in each will keep pace with the other over time (explicitly stated for example in Wulf and McKee (1995) \cite{wulf1995hitting}).
\item That magnetic storage has a practically infinite write cycle lifetime (so thrashing the page file due to insufficient RAM won't cause self-destruction).
\item And therefore we ought to maximise the utilisation of as much scarce and expensive RAM as possible by expending CPU cycles on copying memory around in order to allow programmers to write software for tomorrow's computers by using magnetic storage to ``fake'' tomorrow's memory capacities.
\end{enumerate}

As a result, we have the standard C malloc API (whose design is almost unchanged since the 1970s \cite{seventhunix1979} and which is used by most applications and languages) which takes no account of virtual memory \emph{at all} -- \textbf{each application allocates memory as though it is the sole application in the system}. The kernel then \emph{overrides} the requested memory usage of each process on the basis of a set of statistical assumptions about which pages might actually used by \emph{faking} an environment for the process where it is alone in the system and can have as much memory as it likes -- using the page fault handler to load and save memory page to the page file and polluting all over the processor caches in doing so. Meanwhile you have programmers finding pathological performance in some corner cases in this environment, and writing code which is specifically designed to fool the kernel into believing that memory is being used when it is not, despite that such prefaulting hammers the CPU caches and is fundamentally a waste of CPU time and resources.

This design is stupid, but choosing it was not. There is good reason why it has become so ubiquitous, and I personally cannot think of a better least worst solution given the assumptions above.

\section{Historical trends in growths of capacities and speeds}

\subsection{The increasing non-substitutability of RAM}
\begin{figure}[h]
  \centering
    \includegraphics[width=0.5\textwidth]{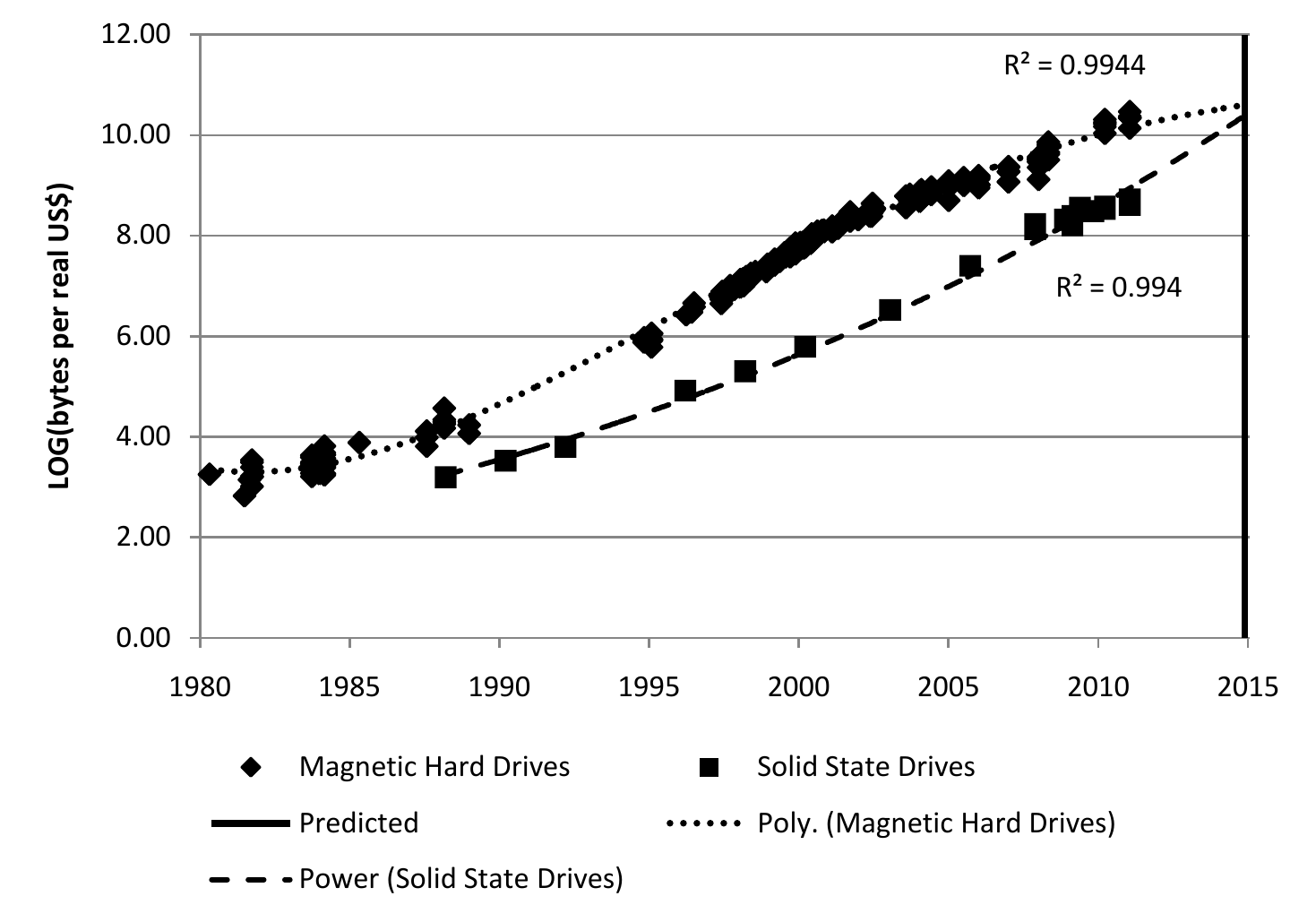}
  \caption{A log plot of bytes available per inflation adjusted US\$ from 1980 to 2010 for conventional magnetic hard drives and flash based solid state disk drives. Magnetic hard drives are clearly coming to the end of their logistic growth curve trajectory, whereas flash drives are still to date undergoing the exponential growth part of their logistic growth curve. Sources: \cite{winch2010, wiki2010b, storsearch2010}.}
  \label{FigSSDsVsHardDrives}
\end{figure}

Unknowable until early 2000's\footnote{It is unknowable because one cannot know when logistic growth will end until well past its point of inflection i.e. when the second derivative goes negative.}, Figure \ref{FigSSDsVsHardDrives} proves the fact that magnetic non-volatile storage is due to become replaced with flash based storage. Unlike magnetic storage whose average random access latency may vary between 8-30ms \cite{tomshardware2010} and which has a variance strongly dependent on the distance between the most recently accessed location and the next location, flash based storage has a flat and uniform random access latency just like RAM. Where DDR3 RAM may have a 10-35ns read/write latency, current flash storage has a read/write latencies of 10-20\textmu s and 200-250\textmu s respectively \cite{desnoyers2010empirical} which is only \textbf{three orders slower} than reading and four orders slower than writing RAM respectively, versus the \textbf{six orders} of difference against magnetic storage. \textbf{What this means is that the relative overhead of page faulted virtual memory on flash based storage becomes \emph{much larger} -- several thousand fold larger -- against overall performance than when based on magnetic storage based swap files}.

Furthermore flash based storage has a limited write cycle lifetime which makes it inherently unsuitable for use as temporary volatile storage. This has a particular consequence, and let me make this clear: \textbf{there is no longer any valid substitute for RAM}.

\subsection{How growth in RAM capacity is going to far outstrip our ability to access it}
\begin{figure}[h]
  \centering
    \includegraphics[width=0.5\textwidth]{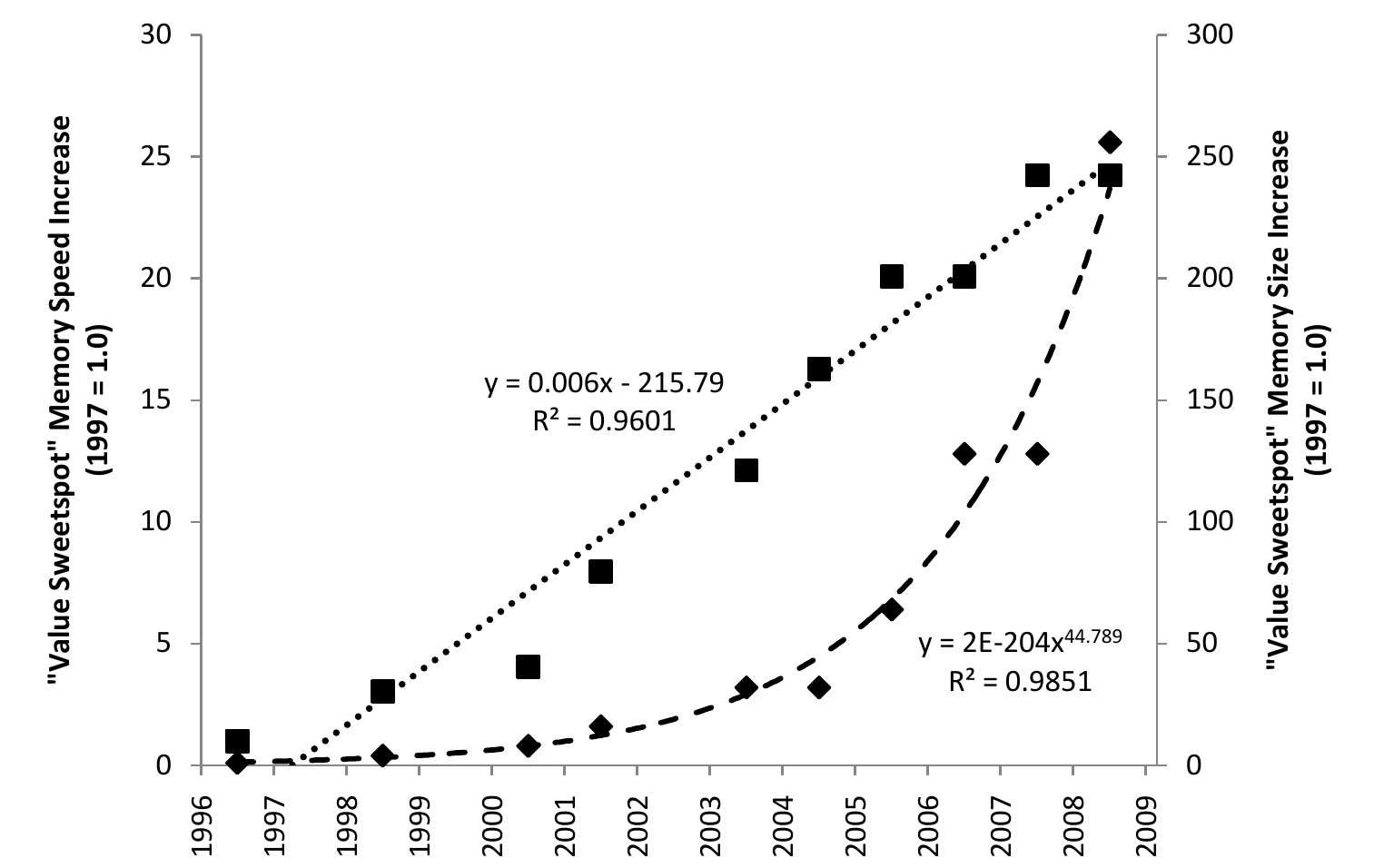}
  \caption{A plot of the relative growths since 1997 of random access memory (RAM) speeds and sizes for the best value (in terms of Mb/US\$) memory sticks then available on the US consumer market from 1997 - 2009, with speed depicted by squares on the left hand axis and with size depicted by diamonds on the right hand axis. The dotted line shows the best fit regression for speed which is linear, and the dashed line shows the best fit for size which is a power regression. Note how that during this period memory capacity outgrew memory speed by a factor of ten. Sources: \cite{jcmit2010} \cite{wiki2010}.}
  \label{FigMemorySizeVsSpeed}
\end{figure}

As Figure \ref{FigMemorySizeVsSpeed} shows, growth in capacity for the ``value sweetspot" in RAM (i.e. those RAM sticks with the most memory for the least price at that time) has outstripped growth in the access speed of the same RAM by a factor of ten in the period 1997-2009, so while we have witnessed an impressive 25x growth in RAM speed we have also witnessed a \textbf{250x} growth in RAM capacity during the same time period. Moreover, growth in capacity is exponential versus a linear growth in access speed, so the differential is only going to dramatically increase still further in the next decade. Put another way, assuming that trends continue and all other things being equal, if it takes a 2009 computer 160ms to access all of its memory at least once, it will take a 2021 computer \textbf{5,070 years} to do the same thing\footnote{No one is claiming that this will actually be the case as the exponential phase of logistic growth will surely have ended before 2021.}.

\subsection{Conclusions}
So let's be really clear:
\begin{enumerate}
\item Non-volatile storage will soon be unsubstitutable for RAM.
\item RAM \textbf{capacity} is no longer scarce.
\item What is scarce, and going to become a LOT more scarce is \textbf{memory access speed}.
\item Therefore, if the growth in RAM capacity over its access speed continues, we are increasingly going to see applications constrained not by insufficient storage, but by insufficiently fast access \emph{to} storage.
\end{enumerate}

The implications of this change are \textbf{profound} for all users of computing technology, but especially for those responsible for the implementations of system memory management. What these trends mean is that \textbf{historical performance bottlenecks are metamorphising into something new}. And that implies, especially given today's relative abundance and lack of substitutability of memory capacity, that page faulted virtual memory needs to be eliminated and replaced with something less latency creating.

\section{Replacing page faulted virtual memory}
\subsection{How much latency does page faulted virtual memory actually introduce into application execution?}

Sadly, I don't have the resources available to me to find out a full and proper answer to this, but I was able to run some feasibility testing. Proper research funding would be most welcome.

\begin{figure}[h]
  \centering
    \includegraphics[width=0.5\textwidth]{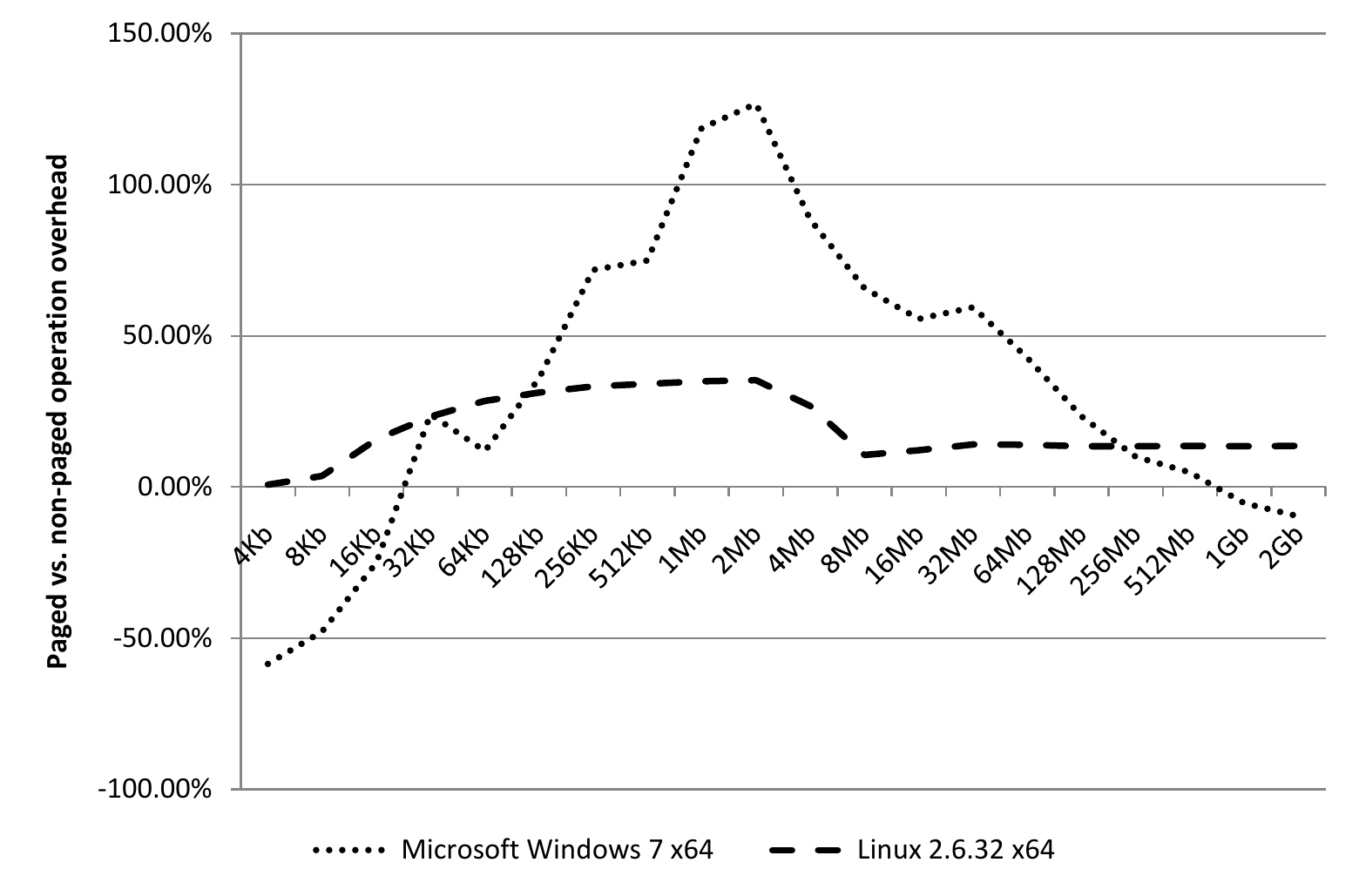}
  \caption{A log-log plot of how much overhead paged virtual memory allocation introduces over non-paged memory allocation according to block size.}
  \label{FigPagedMemoryOverheads}
\end{figure}

Figure \ref{FigPagedMemoryOverheads} shows how much overhead is introduced in a best case scenario by fault driven page allocation versus non-paged allocation for Microsoft Windows 7 x64 and Linux 2.6.32 x64 running on a 2.67Ghz Intel Core 2 Quad processor. The test allocates a block of a given size and writes a single byte in each page constituting that block, then frees the block. On Windows the Address Windowing Extension (AWE) functions \texttt{AllocateUserPhysicalPages()} et al. were used to allocate non-paged memory versus the paged memory returned by \texttt{VirtualAlloc()}, whereas on Linux the special flag \texttt{MAP\_POPULATE} was used to ask the kernel to prefault all pages before returning the newly allocated memory from \texttt{mmap()}. As the API used to perform the allocation and free is completely different on Windows, one would expect partially incommensurate results.

\begin{figure}[h]
  \centering
    \includegraphics[width=0.5\textwidth]{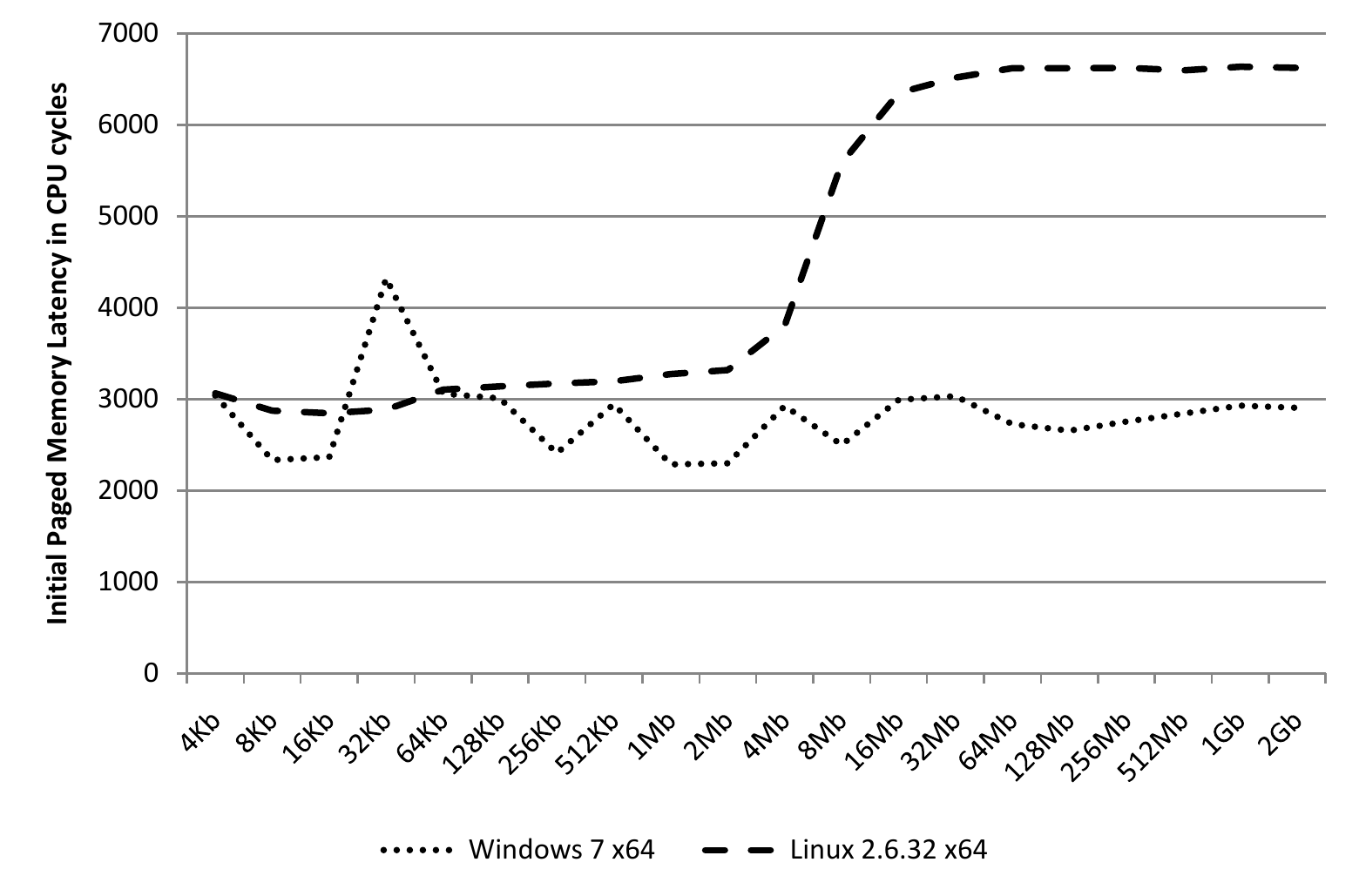}
  \caption{A log-linear plot of how many CPU cycles is consumed per page by the kernel page fault handler when allocating a run of pages according to block size.}
  \label{FigKernelPageFaultHandlerPerformance}
\end{figure}

As one can see, the overhead introduced by fault driven page allocation is substantial with the overhead reaching 125\% on Windows and 36\% on Linux. The overhead rises linearly with the number of pages involved up until approximately 1-2Mb after which it drops dramatically. This makes sense: for each page fault the kernel must perform a series of lookups in order to figure what page should be placed where, so the overhead from the kernel per page fault as shown in Figure \ref{FigKernelPageFaultHandlerPerformance} ought to be approximately constant at 2800 cycles per page for Microsoft Windows 7. There is something wrong with the Linux kernel page fault handler here: it costs 3100 cycles per page up to 2Mb which seems reasonable, however after that it rapidly becomes 6500 cycles per page which suggests a TLB entry size dependency or a lack of easily available free pages and which is made clear in Table \ref{TableMemoryLatencies}. However that isn't the whole story -- obviously enough, each time the kernel executes the page fault handler it must traverse several thousand cycles worth of code and data, thus kicking whatever the application is currently doing out of the CPU's instruction and data caches and therefore necessitating a reload of those caches (i.e. a memory stall) when execution returns.

In other words, fault driven page allocation introduces a certain amount of ongoing CPU cache pollution and therefore raises by several orders not just the latency of first access to a memory page not currently in RAM, but also memory latency in general.

\begin{table}[h]
\caption{Selected page fault allocation latencies for a run of pages on Microsoft Windows and Linux.}
\begin{center}
\begin{tabular}{ r | r r | r r }
& \multicolumn{2}{|c|}{Microsoft Windows} & \multicolumn{2}{|c}{Linux} \\
& \small{Paged} & \small{Non-paged} & \small{Paged} & \small{Non-paged} \\
Size & \small{cycles/page} & \small{cycles/page} & \small{cycles/page} & \small{cycles/page} \\
\hline
16Kb & 2367 & 14.51 & 2847 & 15.83 \\
1Mb & 2286 & 81.37 & 3275 & 14.53 \\
16Mb & 2994 & 216.2 & \textbf{6353} & 113.4 \\
512Mb & 2841 & 229.9 & \textbf{6597} & 115.9 \\
\end{tabular}
\end{center}
\label{TableMemoryLatencies}
\end{table}

The point being made here is that when an application makes frequent changes to its virtual address space layout (i.e. the more memory allocation and deallocation it does), page faulted virtual memory introduces a lot of additional and often hard to predict application execution latency through the entire application as a whole. The overheads shown in Figures \ref{FigPagedMemoryOverheads} and \ref{FigKernelPageFaultHandlerPerformance} represent a best-case scenario where there isn't a lot of code and data being traversed by the application -- in real world code, the additional overhead introduced by pollution of CPU caches by the page fault handler can become sufficiently pathological that some applications deliberately prefault newly allocated memory before usage.

\subsection{What to use instead of page faulted virtual memory}
I propose a very simple replacement of page faulted virtual memory: \emph{a user mode page allocator}. This idea is hardly new: Mach has an external pager mechanism \cite{university1990extending}, V++ employed an external page cache management system \cite{harty1992application}, Nemesis had self-paging \cite{hand1998self} and Azul implements a pauseless GC algorithm \cite{click2005pauseless} which uses a special Linux kernel module to implement user mode page management \cite{azul2011}. However something major has changed just recently -- the fact that we can now use nested page table suppor in commodity PCs to hardware assist user mode page management. As a result, the cost of this proposal is perhaps just a 33-50\% increase in page table walk costs.

The design is simple: one virtualises the MMU tables for each process which requests it (i.e. has a sufficiently new version of its C library). When you call \texttt{malloc}, \texttt{mmap} et al. and new memory is needed, a kernel upcall is used to asynchronously release and request physical memory page frame numbers of the various sizes supported by the hardware and those page frames are mapped by the C library to wherever needed via direct manipulation of its virtualised MMU page tables. When you call \texttt{free}, \texttt{munmap} et al. and free space coalescing determines that a set of pages is no longer needed, these are placed into a \emph{free page cache}. The kernel may occasionally send a signal to the process asking for pages from this cache to be asynchronously or synchronously freed according to a certain severity.

There are three key benefits to this design. The first is that under paged virtual memory when applications request new memory from the kernel, it is not actually allocated right there and then: instead it is allocated \emph{and its contents zeroed} on first access which introduces cache pollution as well as using up lots of memory bandwidth. The user mode page allocator avoids unnecessary pages clears, or when necessary avoids them being performed when the CPU is busy by keeping a cache of free pages around which can be almost instantly mapped to requirements without having to wait for the kernel or for the page to be cleared. Because dirtied pages are still cleaned when they move between processes, data security is retained and no security holes are introduced, but pages are not cleared when they are relocated within the same process thus avoiding unnecessary page clearing or copying.

If this sounds very wasteful of memory pages, remember that \emph{capacities} are increasing exponentially. Even for ``big-iron" applications soon access latencies will be far more important than capacities. One can therefore afford to `throw' memory pages at problems.

The second benefit is that a whole load of improved memory management algorithms become available. For example, right now when extending a \texttt{std::vector<>} in C++ one must allocate new storage and move construct each object from the old storage into the new storage, then destruct the old storage. Under this design one simply keeps some extra address space around after the vector's storage and maps in new pages as necessary -- thus avoiding the over-allocation typical in existing \texttt{std::vector<>} implementations. Another example is that memory can be relocated from or swapped between A and B at a speed invariant to the amount of data by simply remapping the data in question. The list of potential algorithmic improvements goes on for some time, but the final and most important one that I will mention here is that memory management operations can be much more effectively \emph{batched} across multiple cores, thus easily allowing large numbers of sequential allocations and deallocations to be performed concurrently -- something not easily possible with current kernel based designs. As an example of the benefits of batching, consider the creation of a four million item list in C++ which right now requires four million separate \texttt{malloc} calls each of identical size. With a batch malloc API such as N1527 proposed by myself and currently before the ISO C1X standards committee \cite{n1527}, one maps as many free pages as are available for all four million members in one go (asynchronously requesting the shortfall/reloading the free page cache from the kernel) and simply demarcates any headers and footers required which is much more memory bandwidth and cache friendly. While the demarcation is taking place, the kernel can be busy asynchronously returning extra free pages to the process, thus greatly parallelising the whole operation and therefore substantially reducing total memory allocation latency by removing waits on memory.

In case you think a batch malloc API unnecessary, consider that a batched four million item list construction is around 100,000 times faster that at present. And consider that compilers can easily aggregate allocation and frees to single points, thus giving a free speed-up such that C++'s allocation speeds might start approaching that of Java's. And finally consider that most third party memory allocators have provided batch APIs for some time, and that the Perl language found a 18\% reduction in start-up time by adopting a batch malloc API \cite{perl2011}.

The third benefit is implied from the first two, but may be non-obvious: \textbf{memory allocation becomes invariant to the amount allocated}. As mentioned earlier, as capacity usage continues to be exchanged for lower latencies, the amounts of memory allocated and deallocation are going to rise -- a lot. A user mode page allocator removes the overhead associated with larger capacity allocation and deallocation as well as its first time access -- witness the massive drop in latencies shown by Table \ref{TableMemoryLatencies}.

\subsubsection{Testing the feasibility of a user mode page allocator}
\begin{figure}[h]
  \centering
    \includegraphics[width=0.5\textwidth]{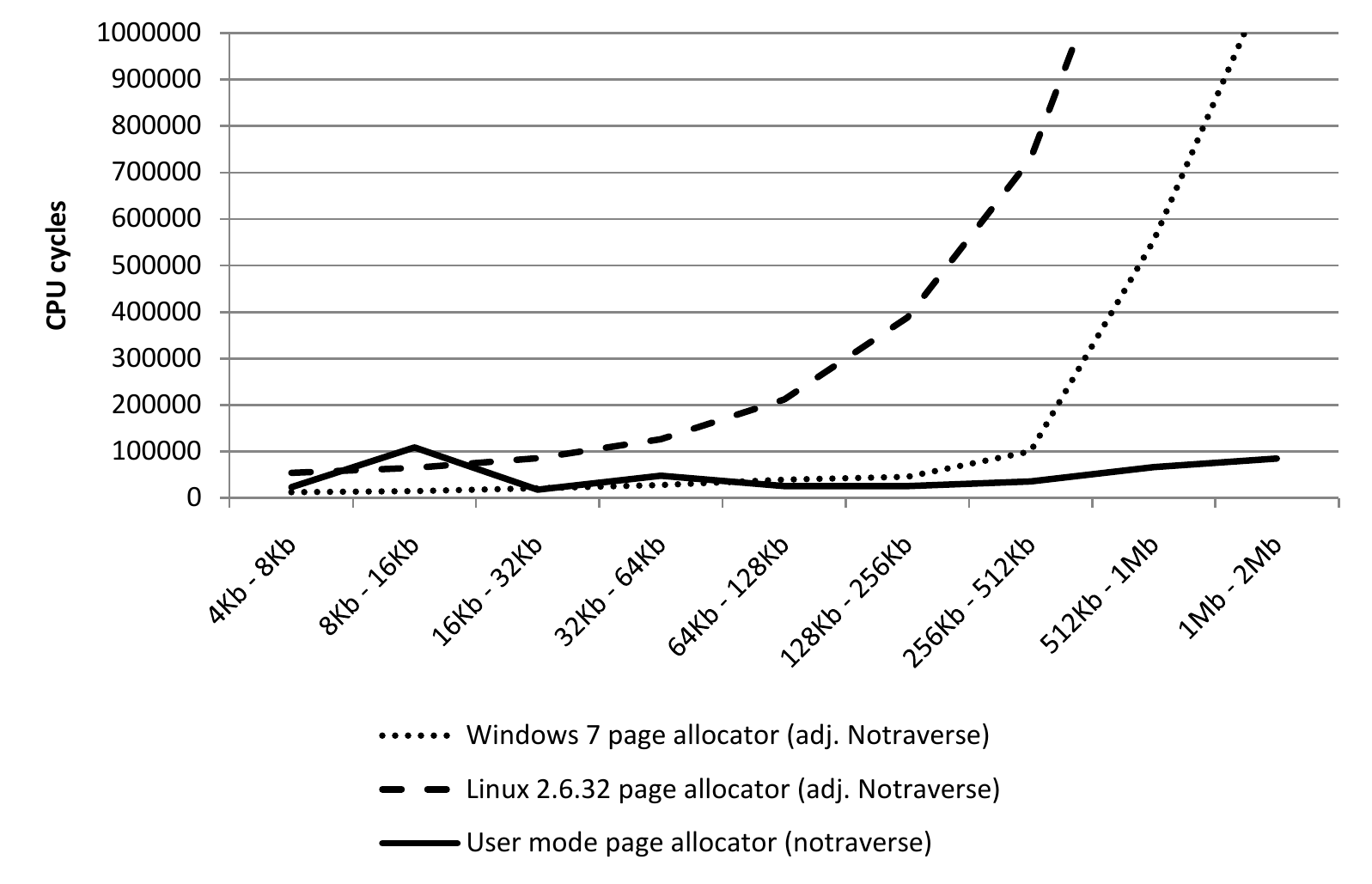}
  \caption{A log-linear plot of how the system and user mode page allocators scale to block sizes between 4Kb and 1Mb. Adjusted no traverse means that single byte write traversal costs were removed to make the results commensurate.}
  \label{FigSystemVsUserModeBreakout}
\end{figure}

To test the effects of a user mode page allocator, a prototype user mode page allocator\footnote{It is open source and can be found at \url{http://github.com/ned14/nedmalloc}.} was developed which abuses the Address Windowing Extensions (AWE) API of Microsoft Windows mentioned earlier in order to effectively provide direct user mode access to the hardware MMU. The use of the verb `abuses' is the proper one: the AWE functions are intended for 32-bit applications to make use of more than 4Gb of RAM, and they were never intended to be used by 64 bit applications in arbitrarily remapping pages around the virtual address space. Hence due API workarounds the prototype user mode page allocator runs (according to my testing) at least 10x slower than it ought to were the API more suitable, and probably more like 40x slower when compared to a memory copy directly into the MMU page tables -- however, its dependency or lack thereof on allocation size in real world applications should remain clear. The results shown by Figure \ref{FigSystemVsUserModeBreakout} speak for themselves.

\begin{figure}[h]
  \centering
    \includegraphics[width=0.5\textwidth]{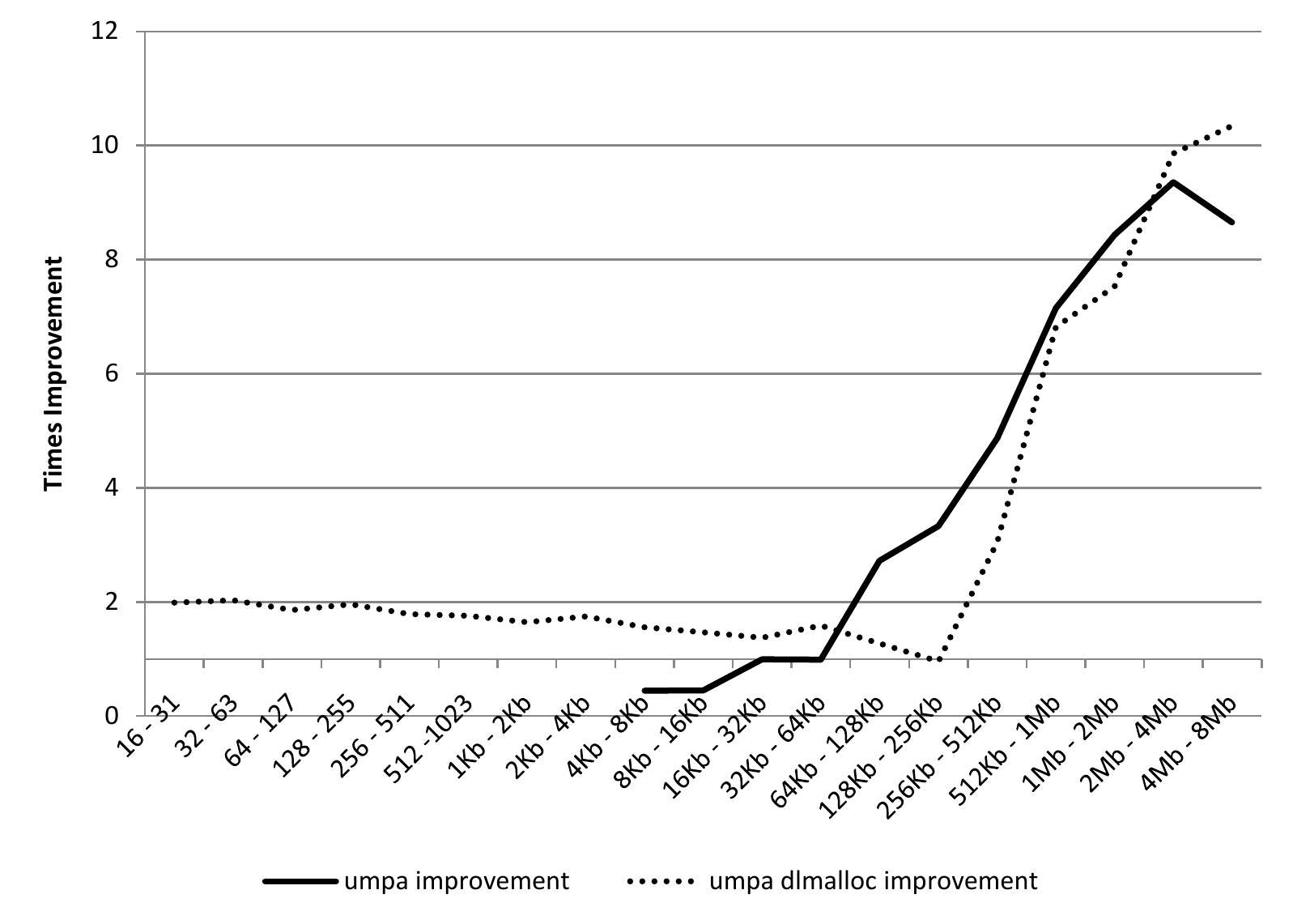}
  \caption{A summary of the performance improvements in the Lea allocator provided by the user mode page allocator (umpa) where 1.0 equals the performance of the system page allocator.}
  \label{FigTimesImprovement}
\end{figure}

Looking at the performance of \texttt{malloc} et al. as provided by the Lea allocator \cite{lea2000memory} under the user mode page allocator (where 1.0 equals the performance of the Lea allocator under the Windows kernel page allocator), Figure \ref{FigTimesImprovement} illuminates an interesting 2x performance gain for very small allocations when running under the user mode page allocator with a slow decline in improvement as one approaches the 128Kb--256Kb range. This is particularly interesting given that the test randomly mixes up very small allocations with very big ones, so why there should be a LOG(allocation size) related speed-up is surprising. I would suggest that the lack of cache pollution introduced by the lack of a page fault handler being called for every previously unaccessed page is most likely to be the cause.

\begin{table}[h]
\caption{Effects of the user mode page allocator on the performance of selected real world applications.}
\begin{center}
\begin{tabular}{ r | r r }
& Peak Memory & \\
Test & Usage & Improvement \\
\hline
Test 1a (G++): & 198Mb & +2.99\% \\
Test 1b (G++): & 217Mb & +1.19\% \\
Test 1c (G++): & 250Mb & +5.68\% \\
Test 1d (G++): & 320Mb & +4.44\% \\
Test 2a (G++): & 410Mb & +3.04\% \\
Test 2b (G++): & 405Mb & +1.20\% \\
Test 2c (G++): & 590Mb & +5.25\% \\
Test 2d (G++): & 623Mb & +3.98\% \\
Test 3a (MS Word): & 119Mb & \textbf{-4.05\%} \\
Test 3b (MS Word): & 108Mb & +0.67\% \\
& & [-0.25\% - +1.27\%], \\
Test 4a (Python): & 6 - 114Mb & avrg. +0.47\% \\
& & [-0.41\% - +1.73\%], \\
Test 4b (Python): & 92 - 870Mb & avrg. +0.35\% \\
Test 5a (Solver): & -- & +1.41\% \\
Test 5b (Solver): & -- & +1.07\% \\
Test 5c (Solver): & -- & +0.58\% \\
\hline
\multicolumn{3}{r}{\emph{Mean = +1.88\%}} \\
\multicolumn{3}{r}{\emph{Median = +1.20\%}} \\
\multicolumn{3}{r}{\emph{Chi-squared probability of independence $p$ = 1.0}} \\
\end{tabular}
\end{center}
\label{TableRealWorldTestResults}
\end{table}

Table \ref{TableRealWorldTestResults} shows the effect of the user mode page allocator in various real world applications binary patched to use the Lea allocator. Clearly the more an application allocates a lot of memory during its execution (G++) rather than working on existing memory (Python), the better the effect.

\section{Conclusion and further work}

Even a highly inefficient user mode page allocator implementation shows impressive scalability and mostly positive effects on existing applications -- even without them being recompiled to take advantage of the much more efficient algorithms made possible by virtualising the MMU for each process.

I would suggest that additional research be performed in this area -- put another way, easy to implement efficiency gains in software could save billions of dollars by delaying development of additional hardware complexity to manage the memory wall. And besides, page file backed memory is not just unnecessary but performance sapping in modern systems.

\acks
The author would like to thank Craig Black, Kim J. Allen and Eric Clark from Applied Research Associates Inc. of Niceville, Florida, USA for their assistance during this research, and to Applied Research Associates Inc. for sponsoring the development of the user mode page allocator used in the research performed for this paper. I would also like to thank Doug Lea of the State University of New York at Oswego, USA; David Dice from Oracle Inc. of California, USA; and Peter Buhr of the University of Waterloo, Canada for their most helpful advice, detailed comments and patience.

\bibliographystyle{unsrtnat}
\softraggedright
\bibliography{MemoryAllocation,ThisPaper}

\end{document}